\documentclass{nime-alternate} %
\usepackage{anonymize}        %
\usepackage[utf8]{inputenc}
\begin{document}

\conferenceinfo{NIME'20,}{July 21-25, 2020, Royal Birmingham Conservatoire, ~~~~~~~~~~~~ Birmingham City University, Birmingham, United Kingdom.}
\title{Sonic Sculpture: Activating Engagement with Head-Mounted Augmented Reality}

\label{key}

\numberofauthors{3}

\author{
  \alignauthor
  \anonymize{Charles Patrick Martin}\\
  \affaddr{\anonymize{Australian National University}}\\
  \affaddr{\anonymize{Canberra, Australia}}\\
  \email{\anonymize{Charles.Martin@anu.edu.au}}
  \alignauthor
  \anonymize{Zeruo Liu, Yichen Wang, Wennan He}\\
  \affaddr{\anonymize{Australian National University and Data 61, CSIRO}}\\
  \affaddr{\anonymize{Canberra, Australia}}\\
  \alignauthor
  \anonymize{Henry Gardner}\titlenote{\anonymize{Henry Gardner is a Visiting Scientist at Data61, CSIRO}}\\
  \affaddr{\anonymize{Australian National University}}\\
  \affaddr{\anonymize{Canberra, Australia}}\\
  \affaddr{\anonymize{Australia}}\\
  \email{\anonymize{Henry.Gardner@anu.edu.au}}
}

\maketitle
\begin{abstract}
This work examines how head-mounted AR can be used to build an interactive sonic landscape to engage with a public sculpture. 
We describe a sonic artwork, ``Listening To Listening'', that has been designed to accompany a real-world sculpture with two prototype interaction schemes. Our artwork is created for the HoloLens platform so that users can have an individual experience in a mixed reality context.
Personal head-mounted AR systems have recently become available and practical for integration into public art projects, however research into sonic sculpture works has yet to account for the affordances of current portable and mainstream AR systems. In this work, we take advantage of the HoloLens' spatial awareness to build sonic spaces that have a precise spatial relationship to a given sculpture and where the sculpture itself is modelled in the augmented scene as an ``invisible hologram''. 
We describe the artistic rationale for our artwork, the design of the two interaction schemes, and the technical and usability feedback that we have obtained from demonstrations during iterative development. 
\end{abstract}




\section{Introduction}

The development of personalised sonic artworks has long been discussed in the computer music community. In the last several years, personal augmented reality (AR) systems have become practical for integration into public art projects, however research into public sonic sculpture works has yet to account for the affordances of portable (untethered)  head-mounted AR systems. %
In particular, systems such as the Microsoft HoloLens, can maintain a real-time spatial map of an environment that contains a physical sculpture and these systems can locate interactive holograms accurately within that spatial environment.
Because of this spatial awareness, it is possible to build sonic spaces that have a precise spatial relationship to a given sculpture. By representing the sculpture itself as an ``invisible hologram", it is possible to exploit the interactivity of AR systems to, for example, play different audio tracks as a user moves near to or touches such an invisible hologram. 

\begin{figure}
  \centering
    \includegraphics[width=\columnwidth]{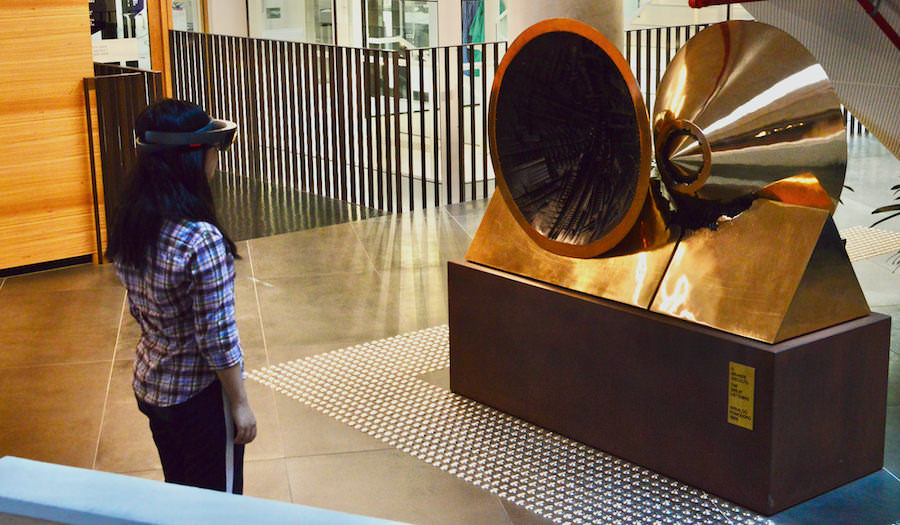}
  \caption{A user engaging with a physical sculpture through our AR sound artwork. A video can be found at: \protect\url{https://doi.org/10.5281/zenodo.3773653}}
  \label{fig:sculpture}
\end{figure}

In this paper, we describe a sonic artwork that has been designed to accompany a real-world sculpture with two contrasting interaction schemes. Our artwork is created for the HoloLens platform so that users can have an individual experience in a mixed-reality context. The HoloLens allows a virtual scene of interactive 3D visual elements and sound sources to be placed around the real sculpture so that users experience both simultaneously. The HoloLens also enables interaction with hand gestures as well as tracking the user's location in the physical world. Our interaction schemes focus on two contrasting methods of interacting with our sound artwork. In the first, users control sounds with a virtual mixer interface placed in front of the sculpture. In the second, the user's location around the sculpture is used to control the sound artwork. Both interaction schemes have the same design goals: 1) to reveal a hidden interactive sound-world; and 2) to activate engagement with the real-world sculpture. Through the process of developing these interactions, and performing a number of demonstrations, we have found that the ways that the interaction schemes achieve these two goals are different, providing lessons for future integration of AR sound art with physical sculpture. 

This work examines how head-mounted AR can be used to build an interactive sonic landscape to engage with a public sculpture. 
Our paper contributes the technique of using an invisible hologram as a means of interaction in augmented reality and two interaction designs which have been assessed through our experiences creating and demonstrating this proof-of-concept system.  %

\subsection{Mixed Reality and Sonic Interaction}

AR involves merging digital and physical worlds, although current definitions vary in terms of required technologies and aspects of reality (visual, sonic, etc) that are considered \cite{Speicher:2019:MR:3290605.3300767}. NIME authors have often merged the real world with a digital sonic world, or used the real world to drive sound works such as in Gaye et al.'s \emph{Sonic City} \cite{GayeHolmquistMaze-SonicCity}. Sonic worlds were combined with physical artworks in Kiefer and Chevalier's work \cite{Kiefer2018}. There are some examples where digital visual, as well as sonic, scenes are layered onto the real world such as \emph{Music Everywhere}, a piano learning system that uses the HoloLens \cite{sglickman2017}, and \emph{Mirror Fugue} that uses projection mapping onto a piano \cite{xxiao2014}. While headphone-only AR has been an effective way to implement AR in gallery and installation settings, such systems have little potential to precisely locate the user around physical objects. By contrast, head-mounted systems such as the HoloLens, provide the opportunity to locate a user, to know what they are looking at, and to enable interaction with virtual objects. So far, this technology has not been applied in sonic art contexts such as sculpture installations, where sounds are generally spatialised with loudspeakers \cite{kiratli2017hive}. In this research, we ask how the HoloLens can be usefully applied to a personal sonic extension of a physical sculpture allowing individuals to craft their own experience.

\subsection{HoloLens for Sonic Art}

\begin{figure}
  \centering
    \includegraphics[width=\columnwidth]{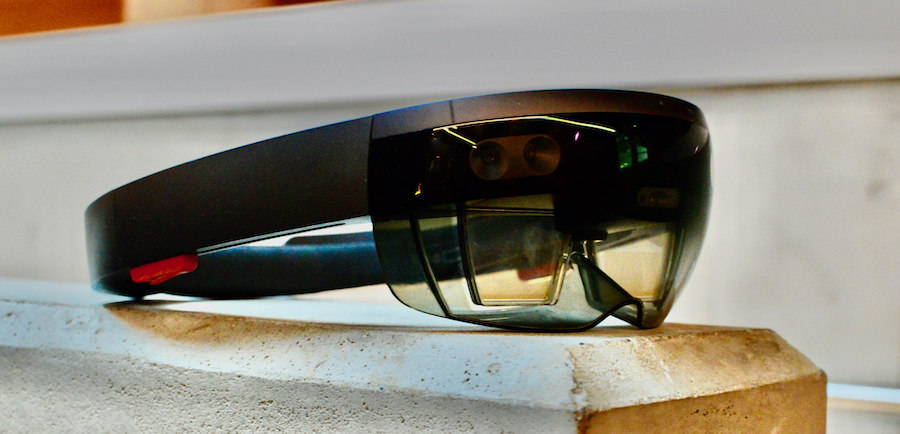}
  \caption{The HoloLens augmented reality wearable computer.}
  \label{fig:hololens}
\end{figure}

The Microsoft HoloLens \cite{msHololens2020} performs a fusion of signals from an inertial sensor, a depth camera and several RGB cameras to locate a user within a model of the world. Once the user's viewpoint has been located, computer graphic holograms and spatialised audio sources can be positioned with respect to their view of that three-dimensional world. These are presented to the user through the transparent display visor and stereo speakers near the user's ears. Although the positioning of the visual holograms is not without error \cite{frantz2018augmenting} such errors are very small relative to macro dimensions of a space within a building and it has been shown that ambisonic sources can be placed sufficiently accurately to enable navigation of a world model by blind users \cite{liu2018augmented}. As such, it is possible to build models of a physical environment that enable a system to understand a sculpture located inside (or outside) of a modelled environment without the use of additional computer vision input for object localisation\cite{frantz2018augmenting}.

Furthermore, by representing the sculpture itself as an ``invisible hologram'', it is possible to exploit the  interactivity of AR holograms in popular development APIs such as Unity \cite{unity2020} to provide the impression that the real sculpture is interactive.
This is what we have set out to achieve in the present project: modelling a physical sculpture, locating that model within a model of the environment, and constructing a three-dimensional, ambisonic sound sculpture in response to that physical sculpture and its environment.  

\begin{figure}
  \centering
  \includegraphics[width=\columnwidth]{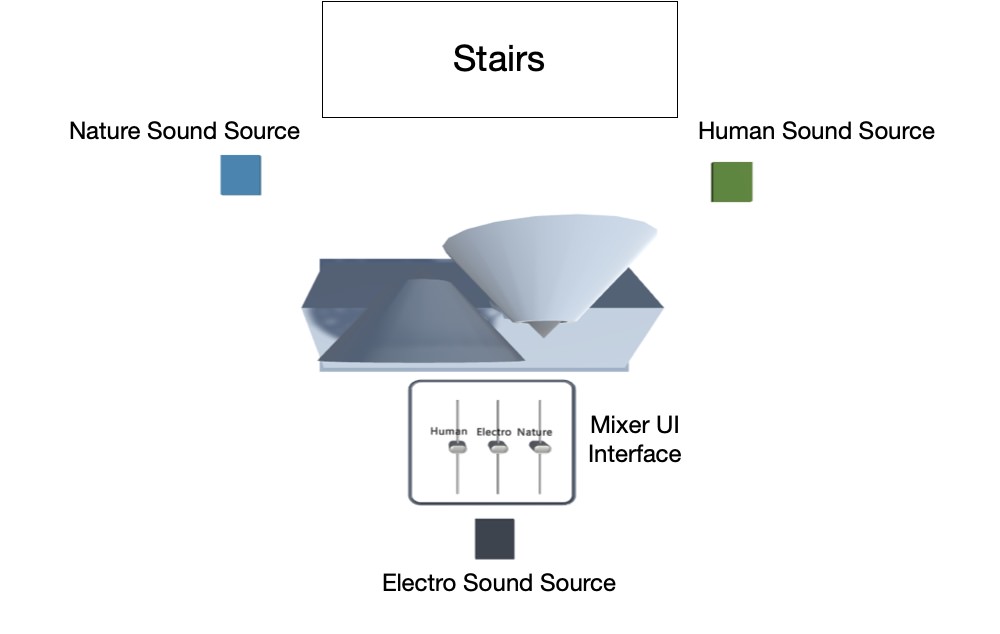}
  \caption{Plan for our installation setup including sculpture location, mixer, and audio source panels. The HoloLens is able to locate the user's position in this space, relative to the sculpture, and present visual elements to them through the display as well as play back spatialised sound from the audio sources.}
  \label{fig:installation-plan}
\end{figure}

\section{System Design}

\emph{Listening To Listening} is an AR sound artwork designed for the
foyer of \anonymize{the Synergy Building at CSIRO's (Australia's
  national research organisation) Black Mountain Site in Canberra,
  Australia}. The sound artwork engages with \emph{Il grande ascolto}
(``The Great Listening'') sculpture by Italian artist Arnaldo
Pomodoro.

The original bronze sculpture depicts two intersecting cones of 80cm
in diameter with shiny bronze exteriors and richly textured interiors.
The sculpture was originally inspired by a radio telescope antenna and
addresses itself to issues of technology and communication
\cite{pomodoro2020}. Curiously, the sculpture has a storied history,
having had one of its cones stolen while located a more prominent
position outside of a different building~\cite{granger_2014}. The
renovated sculpture has been installed underneath a staircase in the
foyer of the present building and is removed from much
pedestrian traffic.

Our sonic AR artwork was designed to accompany and surround \emph{Il
  grand ascolto} and to provide hidden sonic links to technology and
communications, to the open space that it was originally designed for,
and to the present work environment that it now inhabits.

\begin{figure*}
  \centering
    \includegraphics[width=0.32\textwidth]{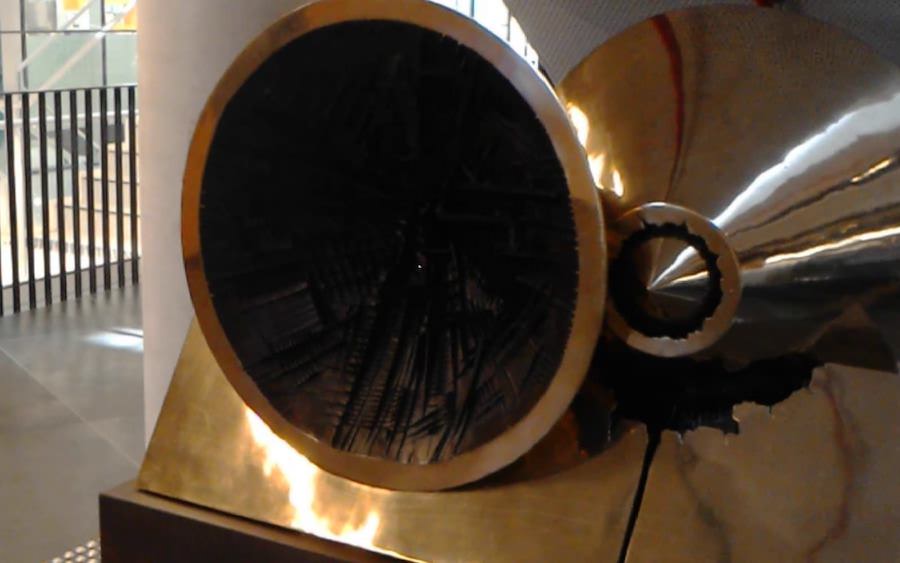}
    \includegraphics[width=0.32\textwidth]{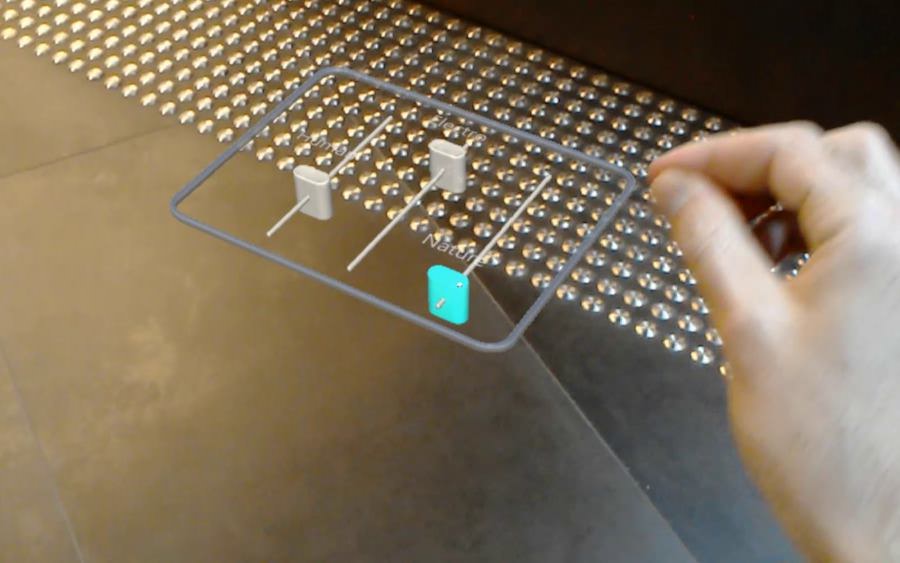}
    \includegraphics[width=0.32\textwidth]{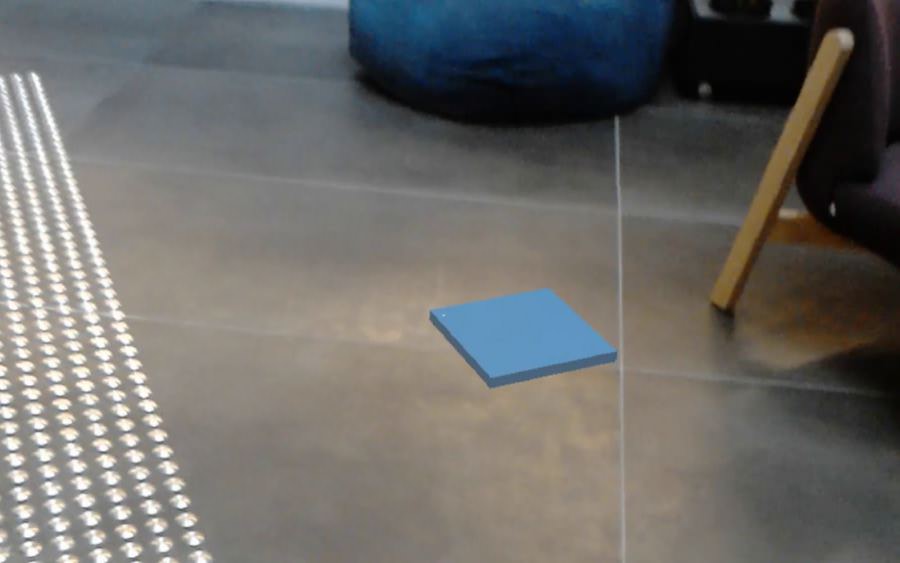}
  \caption{A user's view of the installation: The sculpture (left); mixer interface from interaction A (centre); and floor panels representing the location of audio sources from interaction B.}
  \label{fig:sonic-sculpture-interactions}
\end{figure*}

\begin{figure}
  \centering
    \includegraphics[width=0.75\columnwidth]{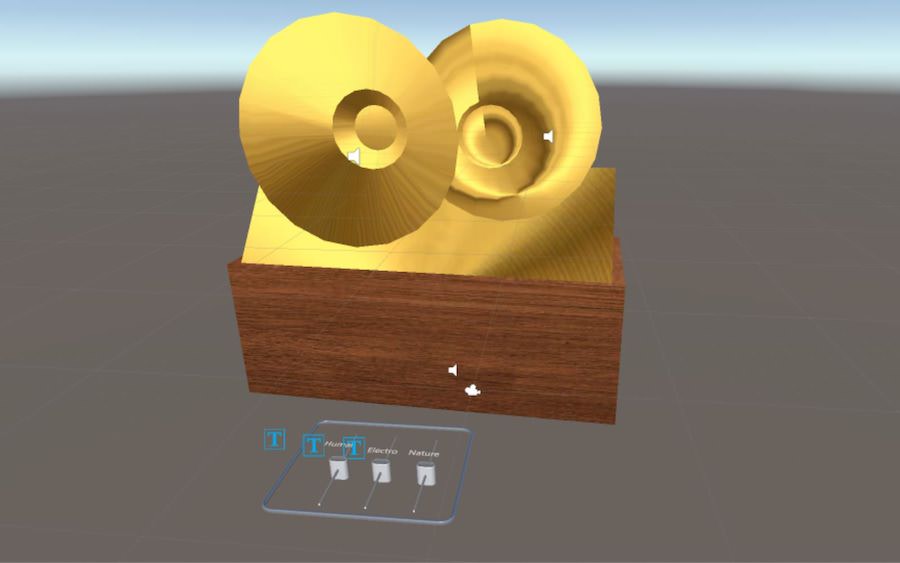} \\
    \includegraphics[width=0.75\columnwidth]{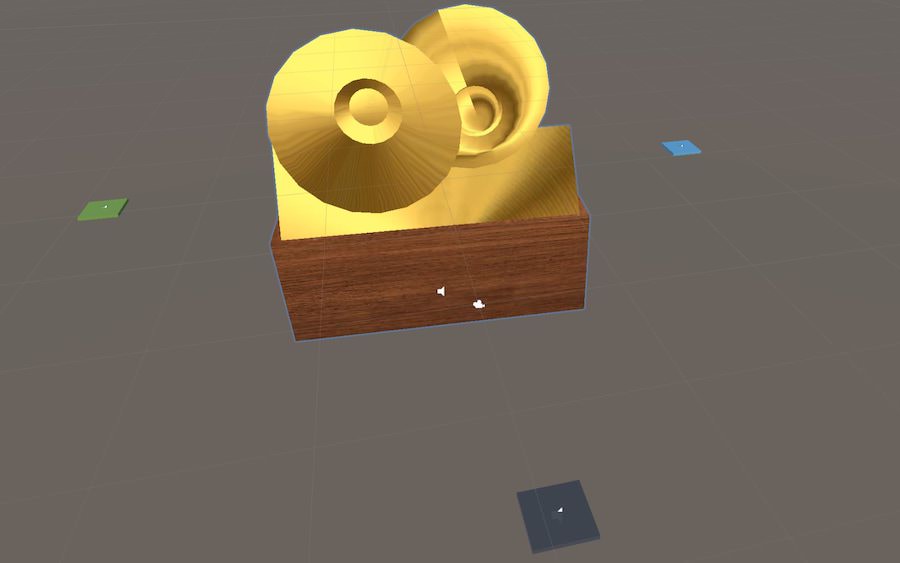}
  \caption{The unity scenes for our two interactions: mixer interface (top), and audio sources represented by panels (bottom).}
  \label{fig:unity-designs}
\end{figure}

\subsection{Sound Artwork}

Our sound artwork is made up of sonic layers of activity where
\emph{Il grande ascolto} is located: 1) the natural environment,
including birdsong and animal activity; 2) the human environment,
including conversation, people, buildings and vehicles; and 3) the
hidden radio-frequency communications of computers and electronics.
Field recordings of all three layers were collected from the
\anonymize{Black Mountain} area where \emph{Il grande ascolto}
sculpture is located, with telephone pickups being used to capture
radio-frequency signals. The recordings were edited and mixed to
produce three six-minute sound works, one for each layer. These three
recordings allow users to listen to three different understandings of
the place and significance of the sculpture, and the sound artwork
encourages users to move and interact with these recordings and engage
more deeply with the idea of the sculpture and its environment.

Listening To Listening was realised with the HoloLens (first generation) AR headset and was programmed in Unity version 2019.2.19. The implementation applied Microsoft's Mixed Reality Toolkit (MRTK) framework \cite{microsoft_2020} which provides a set of features for developing interactive scenes with for the HoloLens. As a first step, a 3D model of the artwork was prepared in Maya. This model was aligned manually with the real sculpture and pinned at the physical location of the artwork using a WorldAnchor~\cite{Unity_2020}, but made invisible during interaction. This means that the HoloLens is able to calculate the physical location of the user with respect to the sculpture and render scenes around it. To explore physical interactions and engagement between the sculpture and sound art work we designed two interaction schemes to explore the three sound layers. The first scheme involved the user being fixed in space but interacting with the sonic layers using UI holograms that were positioned relative to the model of the sculpture. The second scheme used the locative sensing ability of the Hololens as the user moved through the space as guided by holograms positioned relative to the model of the sculpture.

\subsection{Interaction A - UI Mixer}

In this interaction, the user can directly adjust the volume of the three sound layers using UI sliders. We designed three sliders with a frame (resembling a small sound mixer) which is positioned horizontally just above the ground in front of the statue. Each slider controls the level of one sound layer. The audio files are played from Unity AudioSource objects positioned at the sculpture. 
Users can control the three sliders with hand gestures (select and move)~\cite{microsoft_2019}. The audio sliders are all set at the bottom initially and can be set to any combination of volumes. The concept behind this interaction is that the viewer can stand directly in front of the sculpture and explore the three sound layers using direct manipulations. To the user, the sounds appear to emanate from the sculpture so engaging the viewer in both sculpture and sound artwork.

\subsection{Interaction B - Locative Audio Sources}

This interaction uses the user's location around the sculpture to mix
the three sound layers. Each sound layer is played from AudioSource
objects positioned at the vertices of a virtual triangle in the space
surrounding the sculpture's invisible hologram. Three squares were
drawn on the floor to indicate where each AudioSource was centred.

We obtain the Hololens camera position in real time to implement
feedback for HoloLens movement. If the distance between the HoloLens
camera and audio source file is less than 1.5 meters, a custom shader
with white lighting effects will be triggered and the corresponding
square glows subtly as the user approaches. Consequently, as the user
explores the space around the sculpture, they also explore the three
layers of our sound artwork. This interaction represents a natural
control of the audio sources which encourages an inspection of the
artwork from different locations.

\section{Observations}

We have informally trialled our two interaction schemes over a number
of prototyping sessions at the location of the physical sculpture and
in demonstration sessions with project team members and other stakeholders. The observations
that we report here reflect on the design issues and practical
implementation aspects of the two schemes as they might be experienced by real users. In both interaction
schemes, the system's understanding of the location of the user in
relation to the sculpture worked smoothly at closer ranges to the
sculpture itself (e.g., less than 3m away). The mixer UI interaction
scheme was clearly visible. In our early prototypes, we experimented
with where to place these holograms, whether on the sculpture's plinth
or vertically in the air in front of the sculpture. We settled on
placing them horizontally on the floor in front of the sculpture in
order to clearly associate them with the sculpture while not hindering
the user's view of the sculpture. The limited field of view in the
HoloLens meant that it was difficult  to simultaneously
see the sculpture and mixer holograms together (see Figure \ref{fig:sonic-sculpture-interactions}). Controlling the mixer sliders with hand gestures
worked as expected although these gestures were difficult for some
users to operate and were best performed from left-to-right, rather
than in-out. So operation from either side of the sculpture turned out
to be more reliable for this design.

The locative interaction scheme was easier for some users to explore.
The square markers on the ground provided effective guidance as to
where the users might walk. Unlike the mixer interaction, the user was
not able to mix all three sound layers simultaneously, but rather, by
moving from point-to-point, they could mix between two layers at a
time. When users stood right over any one of the squares, the
corresponding audio layer was spatialised directly around them and
sounded very convincing (as if the user had stepped into the sound
world that was being played). This interaction scheme had two
advantages in terms of engagement with the sculpture: Firstly,
experiencing the whole sound world required the user to move around
the sculpture, thus viewing it from multiple sides. Secondly, when the
user looked at the sculpture they could generally see one or another
of the guiding holograms on the floor. This meant that the user was
reminded of other parts of the sound artwork and suggested further
exploration.

\begin{figure}
  \centering
    \includegraphics[width=\columnwidth]{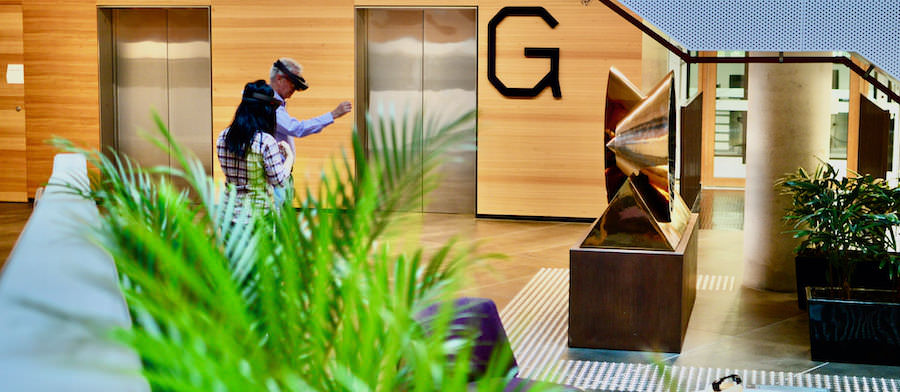}
  \caption{Two users experimenting with UI and locative interaction with our AR sound installation, \emph{Listening To Listening}. Head-mounted AR allows multiple users to have independent experiences simultaneously}
  \label{fig:listening}
\end{figure}

Both of our interactions were sparing in terms of their visual
holograms because our interactions were designed for a space where
there was already effective visual stimuli in the form of the
sculpture itself. Where we were able to use the unique capabilities of
the HoloLens was in terms of locating the UI and sound elements around
the physical sculpture. The mixer interaction provides a similar
interaction to a headphone-based soundart work with a mixer UI
implemented on a mobile device screen. In our case, the HoloLens
co-locates this experience with the sculpture, and the sound appears
to come from the sculpture itself, unifying the existing artwork and
our soundart layer. The locative interaction could potentially be
recreated with physical speakers, but the HoloLens allows this
experience to be individually perceived by different users (see Figure
\ref{fig:listening}) while not disturbing other people in the vicinity
of the sculpture and allows convincing binaural spatialisation of the
sound sources.

\section{Conclusion and Future Work}

We have built an interactive sonic artwork that uses
head-mounted augmented reality to engage with a public sculpture. We
introduced two interaction schemes and we have implemented these
schemes using an ``invisible hologram'', that is, an (unseen) 3D model
of the sculpture itself. While our mixer-interaction scheme involved
longer attention to the front of the sculpture and complete control
over the sound layers, our locative-interaction scheme emphasised
engagement with all sides of the sculpture, and allowed a more
convincing listening experience to individual sound layers. Using the
HoloLens (first gen) was practical for developing a spatialised sonic
artwork, but it did have some limitations such as its restrictive
field of view, limited use of hand gestures, and the limited
possibilities for sound programming in Unity. Future work could
consider using the HoloLens 2 \cite{msHololens2020} (which has a
larger field of view and better gestural recognition) and could also
look at incorporating more sophisticated computer music frameworks
into the HoloLens tool chain.

\subsubsection*{Acknowledgments}

\anonymize{We acknowledge useful discussions with Matt Adcock and Stuart Anderson of Data61, CSIRO and with Ben Swift of ANU.  This research is partially supported by the CSIRO’s Science and Industry Endowment Fund.}

\bibliographystyle{abbrv}
\bibliography{references} 
\end{document}